\documentstyle[prc,aps,epsf,epsfig,floats,twocolumn]{revtex}
\begin{document}
\draft
\twocolumn[\hsize\textwidth\columnwidth\hsize\csname
@twocolumnfalse\endcsname

\title{
%{\small{\tt hep-ph/99mmnnn}}
%{\small {\bf CONFIDENTIAL DRAFT}}
\hfill{\small { FZJ-IKP(TH)-1999-13}}\\[0.2cm]
Strange chiral nucleon form factors}
%Taming Strange Electricity}
%\vspace{1cm} 

\author{
Thomas~R.~Hemmert,
Bastian~Kubis,
Ulf-G. Mei{\ss}ner}

%\vspace{0.5cm}
\address{Forschungszentrum J\"ulich, Institut f\"ur Kernphysik (Th), D-52425 
 J\"ulich, Germany }
\maketitle
%\thispagestyle{empty}

%\vspace{2cm}

\begin{abstract}
We investigate the strange electric and magnetic form factors of
the nucleon in the framework of heavy baryon chiral perturbation theory 
to third order in the chiral expansion. All counterterms can be 
fixed from data. In particular,
the two unknown singlet couplings can be deduced from the parity--violating
electron scattering experiments performed by the SAMPLE and the HAPPEX
collaborations. Within the given uncertainties, our analysis leads
to a small and positive electric strangeness radius, $\langle r_{E,s}^2
\rangle = (0.05 \pm 0.09)\,$fm$^2$. We also deduce the consequences
for the upcoming MAMI A4 experiment.
\end{abstract}

\vspace{1cm}

] 

%%%%%%%%%%%%%%%%%%%%%%%%%%%%%%%%%%%%%%%%%%%%%%%%%%%%%%%%%%%%%%%%%%%%%%%%%%%%%%%%
\section{Introduction}

Recently, the first results from parity--violating electron scattering
experiments, which allow to pin down the so--called strange form factors
of the nucleon, have become available. The SAMPLE collaboration has reported the
first measurement of the strange magnetic moment of the proton~\cite{SAMPLE}. 
To be precise, they give the strange magnetic form factor in units of
nuclear magnetons at a small momentum transfer $Q_S^2=0.1~{\rm GeV}^2$
\begin{eqnarray}
G_{\rm SAMPLE}^{(s)}(Q_S^2)&=&G_M^{(s)} (Q_S^2)\nonumber \\
&=& +0.23\pm 0.37 \pm 0.15\pm 0.19\,.
\end{eqnarray}
The rather sizeable error bars document
the difficulty of such type of experiment.  The HAPPEX collaboration
has chosen a different kinematics which is more sensitive to the
strange electric form factor~\cite{HAPPEX}. Their measurement implies
\begin{eqnarray}
G_{\rm HAPPEX}^{(s)}(Q_H^2)&=&G_E^{(s)}(Q_H^2) + 0.39\;G_M^{(s)}
(Q_H^2) \nonumber \\
&=& 0.023 \pm 0.034 \pm 0.022\pm 0.026\, ,
\end{eqnarray}
with $Q_H^2= 0.48~{\rm GeV}^2$.
There have been many theoretical speculations about the size of the
strange form factors, some of them clearly in conflict with the
data (for a review see ref.\cite{Metal}). 
Here, we wish to analyze these data in the framework of
chiral perturbation theory. It was shown in~\cite{hms} 
that to leading order in the momentum dependence one
can make a parameter--free prediction for the momentum dependence of
the nucleons' strange magnetic (Sachs) form factor based on the chiral symmetry
of QCD solely. The value of the strange magnetic moment, which contains
an unknown low--energy constant, can be deduced from the SAMPLE experiment
using the momentum--dependence derived in~\cite{hms}. Furthermore, the SU(3)
analysis of the octet electromagnetic form factors performed in~\cite{khm}
allows one to pin down the octet component of the strange vector
current. We demonstrate here that to one loop order 
(more precisely: to third order in the chiral expansion) there is only
one new singlet counterterm, whose strength can be determined from 
the value found by HAPPEX. This allows us to give a band for the
strange electric form factor and make a prediction for the MAMI A4
experiment~\cite{MAMI}, which intends to measure  
\begin{eqnarray}
G_{\rm MAMI}^{(s)}(Q_M^2)=G_E^{(s)}(Q_M^2) + 0.22\;G_M^{(s)}
(Q_M^2)\; ,
\end{eqnarray}
with a four-momentum transfer (squared) $Q^2_M=0.23~{\rm GeV}^2$
of approximately half the HAPPEX value.

\medskip
\noindent
The strangeness vector current of the nucleon is defined as
\begin{eqnarray}\label{svc}
\langle N|\;\bar{s}\;\gamma_\mu\; s\;|N \rangle
&=& \langle N|\;\bar{q}\;\gamma_\mu\;
(\lambda^0/3-\lambda^8/\sqrt{3}) \; q\;| N \rangle \nonumber \\
                                 &=&(1/3)J_{\mu}^0- (1/\sqrt{3})J_{\mu}^8 \; ,
\end{eqnarray}
with $q=(u,d,s)$ denoting the triplet of the light quark fields and
$\lambda^0 = I\; (\lambda^a)$ 
the unit (the $a=8$ Gell--Mann) SU(3) matrix. Chiral perturbation theory
(CHPT) is a precise tool to investigate such type of
low--energy properties of the nucleon~\cite{ect}. 
In the past few years, however, it was believed
that due to the appearance of higher order  local contact terms with
undetermined coefficients, CHPT can not be used to make any
prediction for the strange magnetic moment or the strange electric
form factor~\cite{mus}.
However, with the advent of the first SAMPLE and HAPPEX data and renewed theoretical 
effort the situation has now changed. It could be shown that 
to third order in small momenta and/or meson mass
insertions (we collectively denote these expansion parameters by $p$),
there appear  only four low--energy constants (LECs)  
in the octet (note that only two combinations of these are relevant here)
and two in the singlet current. While the former (two combinations)
can be fixed from the
isoscalar anomalous magnetic moment and charge radius of the nucleon, 
the latter two can now be
deduced from the pioneering SAMPLE and HAPPEX results.\footnote{As a 
cautionary remark we
mention already here that the momentum transfer in the HAPPEX
experiment might be too large to trust the third order CHPT treatment. However,
at the moment we consider the experimental uncertainties associated with the 
SAMPLE and HAPPEX input into our calculation to be larger than the 
theoretical uncertainty of truncating the calculation at ${\cal O}(p^3)$.
Ultimately, the situation can be improved by performing the calculation to next order,
smaller experimental error bars and utilizing new data at lower $q^2$, which should 
become available within the next few years.}

%%%%%%%%%%%%%%%%%%%%%%%%%%%%%%%%%%%%%%%%%%%%%%%%%%%%%%%%%%%%%%%%%%%%%%%%%%%%%%%%%%
\section{Theoretical framework}

In order to obtain the strange electric and magnetic (Sachs) form factors we are 
calculating the singlet and the octet current matrix element
of the nucleon to ${\cal O}(p^3)$ in SU(3) HBCHPT, see eq.(\ref{svc}),
in the Breit frame (following refs.\cite{BKKM,BFHM})
%\begin{eqnarray}{\label{current}}
%\langle N(p') |J_\mu^{(0,8)}| N(p)\rangle
%&=& \frac{1}{{\cal N}_{i}{\cal N}_{f}} {\bar u(p')} \, P^{+}_{v} 
%\biggl[G_{E}^{(0,8)} (Q^{2}) v_\mu \nonumber \\
%+ \, \frac{1}{m} \,  G_{M}^{(0,8)} && 
%\!\!\!\!\!\!\! (Q^{2})\, [ S_\mu,S_\nu ]
%   q^{\nu} \biggr] \, P^{+}_{v}\, u(p)~, 
%\end{eqnarray} 
\begin{eqnarray}{\label{current}}
J_\mu^{(0,8)}
&=& \frac{1}{{\cal N}_{i}{\cal N}_{f}} {\bar u(p')} \, P^{+}_{v} 
\biggl[G_{E}^{(0,8)} (Q^{2}) v_\mu \nonumber \\
&& + \, \frac{1}{m} \,  G_{M}^{(0,8)} (Q^{2})\, [ S_\mu,S_\nu ]
   q^{\nu} \biggr] \, P^{+}_{v}\, u(p)~, 
\end{eqnarray} 
with %%$|N(p)\rangle$ a nucleon state of four--momentum $p$, 
\begin{equation}
q_\mu = (p' -p)_\mu~, \quad Q^2 = -q^2~, \quad 
{\cal N} =\sqrt{\frac{E+m}{2m}}~,
\end{equation}
$P_v^+$ being  a positive--velocity  projection operator and $m$
is the nucleon mass. For a
more detailed discussion of this expression and the relation
to the standard Dirac and Pauli form factors, see e.g.~\cite{BFHM}.
From Eq.(\ref{current}) one can then reconstruct the  strangeness form factors
as follows
\begin{eqnarray}
G_{E/M}^{(s)}\left( Q^{2}\right)  &=& \frac{1}{3} G_{E/M}^{(0)}\left( Q^{2}\right)
- \frac{1}{\sqrt{3}} G_{E/M}^{(8)}\left( Q^{2}\right) .  
\end{eqnarray}
These form factors admit a Taylor expansion around $Q^2=0$,
\begin{eqnarray}\label{taylor}
G_{E/M}^{(s)}\left( Q^{2}\right)  &=& G_{E/M}^{(s)} (0) - \frac{1}{6}
\langle r^2_{E/M,s} \rangle \, Q^2 + {\cal O}(Q^4)~,
\end{eqnarray}
in terms of the strange electric/magnetic radii 
\begin{eqnarray}\label{defrad}
\langle r^2_{E/M,s} \rangle = -6 \frac{dG_{E/M}^{(s)} ( Q^{2})}{dQ^2}\biggl|_{Q^2=0}~.
\end{eqnarray}
Note that one does not divide through the normalization of the
respective form factor (even if it is non--vanishing) as it is usually
done in case of the standard electromagnetic Sachs form factors. For
that reason, one sometimes also introduces the slope parameters
$(\rho^{(s)}_{E/M})^2 = \langle r^2_{E/M,s} \rangle / 6$~\cite{mus}.

\medskip

\noindent
We give now the relevant HBCHPT Lagrangians needed for the
calculation.  Throughout, we work in the isospim limit $m_u = m_d$.
We utilize  the covariant derivative acting on the baryon field $B$ in the
fundamental representation
\begin{eqnarray}
D_\mu B&=& \partial_\mu B +[\Gamma_\mu,B] -i \langle v_{\mu}^{(0)} 
\rangle B \nonumber \\
\Gamma_\mu&=&\frac{1}{2}\left[u^\dagger,\partial_\mu u\right]-\frac{i}{2}u^\dagger
             v_{\mu}^{(8)} u-\frac{i}{2}u \,v_{\mu}^{(8)} u^\dagger +\dots \; , 
\end{eqnarray}
respectively the chiral vierbein
\begin{eqnarray}
u_\mu&=&i\;u^\dagger\left(\partial_\mu U-i v_{\mu}^{(8)} U+i U 
               v_{\mu}^{(8)}\right) u^\dagger +\dots \, ,
\end{eqnarray}
where the quantity $v_{\mu}^{(8)}\,[v_{\mu}^{(0)}]$ corresponds to an external octet
[singlet] vector source and
$\langle \ldots \rangle$ denotes the trace in flavor space.
The relevant SU(3) HBCHPT Lagrangians then read (we do not show the
terms which are
%stem from the $1/m$ expansion of the lower order
%Lagrangians and thus have fixed coefficients and also not the ones 
only needed for wave function renormalization)  
\begin{eqnarray}
{\cal L}_{MB}^{(1)}&=&\langle \bar{B}\;i v\cdot D\;B\rangle
                    +  D\; \langle \bar{B}\;S^\mu\{u_\mu,B\} \rangle 
               \nonumber \\
            & & +F \;\langle \bar{B}\;S^\mu[u_\mu,B]\rangle ,\\
{\cal L}_{MB}^{(2)}&=& -\frac{i (1+b^F)}{4 m}\; \langle \bar{B}
               \left[S^\mu,S^\nu\right] [f_{+\mu\nu}^{(8)},B]\rangle
              \nonumber \\
            & &-\frac{i b^D}{4 m}
               \;\langle \bar{B}\left[S^\mu,S^\nu\right] \{f_{+\mu\nu}^{(8)},B\}
               \rangle \nonumber \\
            & & -\frac{i (1+b_{0}) }{4 m}\;
              \langle \bar{B}\left[S^\mu,S^\nu\right] B\rangle 
              \; 2 \langle v_{\mu\nu}^{(0)} \rangle \nonumber \\
            & & -\frac{1}{2m} \;\langle \bar{B}
                    [D_\mu,[D^\mu,B]]\rangle \nonumber \\
            & & +\frac{1}{2m} \;\langle \bar{B}
                    [v\cdot D,[v\cdot D,B]]\rangle~, \\
{\cal L}_{MB}^{(3)}&=& 
-{d^{101} \over (4\pi F_\phi)^2}\langle 
\bar{B} [[ v^\mu D^\nu, f_{+\mu\nu}^{(8)}], B] \rangle \nonumber \\
&&
-{d^{102} \over (4\pi F_\phi)^2}  \langle 
\bar{B} \{[ v^\mu D^\nu, f_{+\mu\nu}^{(8)}], B\}  \rangle \nonumber \\
&&  - {d^{102}_0 \over (4\pi F_\phi)^2}
\langle  \bar{B} B \rangle \langle [ v^\mu \partial^\nu , 2
                    v_{\mu\nu}^{(0)}] \rangle \nonumber \\
&&+ \frac{1}{2m}\; \langle \bar{B} \bigl( \gamma_0 {\cal B}^{(2)\dagger}
                    \gamma_0 {\cal B}^{(1)} +\gamma_0 {\cal B}^{(1)\dagger} 
                    \gamma_0 {\cal B}^{(2)} \bigr) B\rangle \nonumber
                    \\
&& -\frac{1}{4m^2}\langle \bar{B} \gamma_0 {\cal B}^{(1)\dagger}\gamma_0
(iv\cdot D) {\cal B}^{(1)} B \rangle + \ldots~,
\nonumber\\ &&
\label{eq:lag}
\end{eqnarray}
with
\begin{eqnarray}
f_{+\mu\nu}^{(8)}&=&u^\dagger\left(\partial_\mu v_{\nu}^{(8)}-
\partial_\nu v_{\mu}^{(8)}\right) u+u\left(\partial_\mu v_{\nu}^{(8)}-
\partial_\nu v_{\mu}^{(8)}\right) u^\dagger  \nonumber \\
v_{\mu\nu}^{(0)}&=&\partial_\mu v_{\nu}^{(0)}-\partial_\nu
v_{\mu}^{(0)} \; ,
\end{eqnarray}
and the matrices ${\cal B}^{(1,2)}$ encode the information concerning
the $1/m$ corrections due to transitions between the light and heavy
components~\cite{BKKM}. 
Their explicit form for the SU(3) case can be found in ref.\cite{guido}.
Furthermore, $F_\phi =(F_\pi+F_k)/2 \simeq 100\,$MeV is the average 
pseudoscalar decay constant. We use
this value because the difference between the pion and the kaon decay
constants only shows up at higher order. 
For the conventional axial meson--baryon couplings we will use $F =
0.5,\;D=0.75$.\footnote{Note that the symbol $D$ is used for the
  covariant derivative and for one of the axial coupling constants. From the
  context it is, however, always obvious which one is meant.}  
The LECs $d^{101}$, $d^{102}$ have already been determined
in~\cite{khm} from the electric radii of the proton and the neutron.
In contrast to ref.\cite{hms,mus} we separate the anomalous and
non--anomalous contributions to the magnetic moments, utilizing the 
path integral formalism of~\cite{BKKM}.
To make contact with the notation used in ref.\cite{hms}, we notice
that the corresponding dimension two LECs $b^{D/F}_{6b}$ and $b_{bc}$
are related to the ones given above in the following way:
\begin{equation}
b_{6b}^F := 1+ b^F~, \quad
b_{6b}^D := b^D~, \quad
b_{6c} := 3\,(1 +b_0)~.
\end{equation}
The first two of these are nothing but the two SU(3) parameters
originally introduced by Coleman and Glashow~\cite{CG} to derive
relations between the magnetic moments of the octet baryons.
The dimension two LECs are finite numbers since loop corrections only start
at third order.
\medskip

\noindent With these Lagrangians, we are now in the position to
evaluate the strange form factors. Consider first the singlet
contributions. To third order in the chiral expansion, these are given
entirely in terms of tree graphs and therefore take the very simple forms
\begin{eqnarray}\label{siff}
G_{E}^{(0)} (Q^2) &=& 3 \biggl(1 + \frac{1}{\left( 4 \pi F_\phi \right)^{2}} \,
2 \,d_{0}^{102}\, Q^2 
-\frac{1}{4m^2}\,b_0\, Q^2\biggr)~,
\nonumber \\
G_{M}^{(0)} (Q^2) &=&3(1+ b_0) = G_{M}^{(0)} (0) := 3+ \kappa_N^{(0)}~,
\end{eqnarray}
with  $\kappa_N^{(0)}$ the singlet nucleon  anomalous magnetic moment. 
Since there are no loop contributions to this order, the LEC
$d_{0}^{102}$ is finite and scale--independent. Furthermore,
the last term in the electric form factor is the singlet Foldy
term, i.e. we can rewrite the expression for $G_{E}^{(0)}$ as
\begin{equation}\label{rad0}
G_{E}^{(0)} (Q^2) = 3\,\biggl(1+ \biggl[\frac{2d_0^{102}}{(4\pi F_\phi)^2} 
- \frac{\kappa_N^{(0)}}{12m^2} \biggr]  \, Q^2\biggr) ~,
\end{equation}
where the term in the square brackets is (up to a factor) the singlet
electric radius squared, see eq.(\ref{taylor}).
Such a structure is of course familiar from the expression for the
neutron charge radius where the dominant contribution to the 
radius comes indeed from the Foldy term. The precise splitting for the
strange electric radius will be discussed below.
The normalization of $G_{E,M}^{(0)}$ is related to our normalization of the 
singlet current. It is defined as in~\cite{hms} with respect to the 
(valence) quark number
and not the baryon number as often done, see e.g.~\cite{mus}.
There are no loop corrections to the singlet electric charge because
the baryon number current is conserved. There are also no loop
contributions to the strange
electric radius since $v_\mu^{(0)}$ does not couple to the meson
cloud and all graphs with couplings to the nucleon are
momentum--independent to third order. This will change at ${\cal O}(p^4)$.
The singlet magnetic form factor in eq.(\ref{siff}) behaves similarly to the
isoscalar magnetic form factor in SU(2), i.e. to third order it
is entirely given in terms of a dimension two contact term with no
momentum dependence. 

\medskip

\noindent We now discuss $J_\mu^8$. The corresponding octet components
are of course implicitly contained  in ref.\cite{khm} since the electromagnetic
current is an appropriate combination of triplet and octet components. Indeed,
to this order the octet form factor can be calculated from the sum of
the physical proton and neutron form factors and at this order happens to be equal 
to the isoscalar electromagnetic form factor of the nucleon,
%is identical (up to a factor) to the isoscalar $(I=0)$ nucleon form factor,
\begin{eqnarray}\label{octiso}
%G_{E/M}^{I=0}\left( Q^{2}\right)  &=& \left( 1/\sqrt{3}\right) G^{(8)}_{E/M}
%\left(Q^{2}\right) \nonumber \\
G_{E/M}^{(8)}\left( Q^{2}\right)  &=&  \sqrt{3} \;
\left[ G_{E/M}^{p}\left( Q^{2}\right) +G_{E/M}^{n}\left( Q^{2}\right)\right]
\nonumber \\
&=& \sqrt{3}\; G_{E/M}^{I=0}\left( Q^{2}\right) + {\cal O}(p^4)~. 
\end{eqnarray}
After standard renormalization  to take care of the divergences as 
detailed in~\cite{guido}, the corresponding octet electric form factor can 
thus be written as
\begin{eqnarray}\label{GEoct}
G_{E}^{(8)}\left( Q^{2}\right)  &=& 
\sqrt{3} +\frac{\sqrt{3}}{\left( 4\pi F_{ \phi}\right) ^{2}}
\nonumber \\
\times \biggl\{  \biggl[ && \!\!\!\!\!
\frac{1}{12}+\frac{85}{108}D^{2}-\frac{17}{18}DF+\frac{17}{12}F^{2}
\nonumber \\
+
\biggl( \frac{1}{2} &+& \!\!
\frac{5}{6}\biggl( \frac{5}{3} D^{2}- 2DF+3F^{2}\biggr) \biggr) 
\ln \left( \frac{M_{K}}{ \mu}\right) 
\biggr] Q^{2} 
\nonumber \\
+\biggl[ \biggl( && \!\! \frac{5}{3}D^{2} -
2DF+3F^{2}\biggr) \left(2M_{K}^{2}+\frac{5}{4}Q^{2}\right) \nonumber \\ 
&+& 3 \left( M_{K}^{2}+\frac{1}{4}Q^{2}\right)  
\biggr] I_{E}^{K}\left( Q^{2}\right)  
\nonumber  \\
&+& 2\left( d^{101} (\mu ) - \frac{1}{3}d^{102} (\mu )  \right) Q^{2}
\biggr\}\nonumber  \\   
&+& \frac{\sqrt{3}}{4m^{2}}\left(\frac{1}{3}b^{D} - b^{F}  \right) Q^{2} 
\end{eqnarray}
where
\begin{eqnarray}
I_{E}^{K}\left( Q^{2}\right)  &=&\frac{1}{3} \int\limits_{0}^{1}dx
\ln \left( 1 + x ( 1-x) \frac{Q^2}{M_{K}^{2}}\right)~,  
\end{eqnarray}
with  $M_K = 494\,$MeV the kaon mass
and $\mu$ is the scale of dimensional regularization. Throughout, we
set $\mu =1\,$GeV and the scale--dependent LECs are also given at that
scale. They can be evaluated for any other scale making use of the
$\beta$--functions given in ref.\cite{guido}. The corresponding octet
radius can be written as
\begin{eqnarray}\label{rad8}
\langle r_{8,E}^2 \rangle &=& -\frac{\sqrt{3} (b^D- 3b^F)}{2m^2}
- \frac{1}{32\sqrt{3}\pi^2 F_\phi^2} \nonumber\\
&\times& \biggl( 7(5D^2-6FD+9F^2) + 9 \nonumber\\
&+&  72 \, d^{102}(\mu ) -24 \, d^{101} (\mu
) \nonumber \\
&+& 2(5(D^2-6DF+9F^2)+9)\ln\frac{M_K}{\lambda}  \biggr)~,
\end{eqnarray}
using
\begin{equation}
I_{E}^{K}\left( Q^{2}\right) = \frac{1}{18}\frac{Q^2}{M_K^2} + 
{\cal O}\left(\frac{Q^4}{M_K^4}\right)~.
\end{equation}
Similarly, the magnetic octet form factor takes the form
\begin{eqnarray}\label{GM8}
G_{M}^{(8)}\left( Q^{2}\right)  &=&
\sqrt{3} \biggl( 1-\frac{1}{3}b^{D}+b^{F} \biggr)
\nonumber \\
&-& \frac{\sqrt{3}m}{16\pi F_{ \phi}^{2}}
\biggl\{ 
\left( \frac{5}{3} D^{2}-2DF+3F^{2}\right) \nonumber \\
&\times& \left[
M_{K}+\left( M_{K}^{2}+\frac{1}{4}Q^{2}\right)
I_{M}^{K}\left( Q^{2}\right)
\right]\biggr\}
\end{eqnarray}
where
\begin{equation}
I_{M}^{K}\left( Q^{2}\right)  =\int\limits_{0}^{1}
\frac{dx}{\sqrt{M_K^2 + x\left( 1-x\right) Q^{2}}}~.
\end{equation}
To further disentangle the momentum dependence of this form factor,
we bring it into the following compact form,
\begin{eqnarray}\label{GEoct2}
G_M^{(8)} (Q^2) &=& \sqrt{3} + \kappa^{(8)} - \frac{2\sqrt{3}}{3} \frac{
 \pi m M_K}{(4\pi F_\phi)^2} \nonumber\\
&\times& (5D^2 - 6DF + 9F^2) \, f(Q^2)~,
\end{eqnarray}
with the octet anomalous magnetic moment 
\begin{eqnarray}
\kappa^{(8)} &=& \sqrt{3}\; \biggl( b_F -\frac{1}{3}b_D \nonumber \\  
&& -\frac{m M_K}{ 24 \pi F_\phi^2}(5D^2 - 6DF + 9F^2)\biggr)~,
\end{eqnarray}
and the function $f(Q^2)$  given in ref.\cite{hms}. To this order,
we have $\kappa^{(8)} = \sqrt{3} (\kappa_p+\kappa_n)$ due to
eq.(\ref{octiso}). This relation
is trivially fulfilled if one fits the LECs $b_D$ and $b_F$ to the 
neutron and proton magnetic moments using the third order
formula. In fact, the form of $G_M^{(8)}$ as given in eq.(\ref{GM8})
and  eq.(\ref{GEoct2}) differs by the loop contribution to the
magnetic moments. This difference is, however, of higher order.
In what follows, we will work with the form of the octet form factor
given in eq.(\ref{GM8}). We remark that to this order in
the chiral expansion, the momentum dependence of the magnetic octet
form factor completely determines the one of the strange magnetic form
factor.

\medskip

\noindent
Putting pieces together, the strange electric form factor of the nucleon
takes the form 
\begin{eqnarray}\label{GEs}
G_{E}^{(s)}\left( Q^{2}\right)  &=& \frac{1}{\left( 4\pi F_{ \phi}\right) ^{2}}
\nonumber \\
\times \biggl\{ - \biggl[ && \!\!\!\!\!
\frac{1}{12}+\frac{85}{108}D^{2}-\frac{17}{18}DF+\frac{17}{12}F^{2}
\nonumber \\
+
\biggl( \frac{1}{2} &+& \!\!
\frac{5}{6}\biggl( \frac{5}{3} D^{2}- 2DF+3F^{2}\biggr) \biggr) 
\ln \left( \frac{M_{K}}{ \mu}\right) 
\biggr] Q^{2} 
\nonumber \\
-\biggl[ \biggl( && \!\! \frac{5}{3}D^{2} -
2DF+3F^{2}\biggr) \left(2M_{K}^{2}+\frac{5}{4}Q^{2}\right) \nonumber \\ 
&+& 3 \left( M_{K}^{2}+\frac{1}{4}Q^{2}\right)  
\biggr] I_{E}^{K}\left( Q^{2}\right)  
\nonumber \\
&-& 2\left( d^{101}(\mu )-\frac{1}{3}d^{102}(\mu ) - d_{0}^{102} \right) Q^{2}
\biggr\}\nonumber  \\  
&-& \frac{1}{4m^{2}}\left(b_{0}+\frac{1}{3}b^{D} - b^{F}  
\right) Q^{2}~. 
\end{eqnarray}
The strange electric radius can readily be deduced from
eq.(\ref{GEs}), singlet and octet radi given before, see
eqs.(\ref{rad0},\ref{rad8}), via
\begin{equation}
\langle r_{E,s}^2 \rangle = \frac{1}{3} \, \langle r_{E,0}^2 \rangle
- \frac{1}{\sqrt{3}} \, \langle r_{E,8}^2 \rangle~
\end{equation}
In this formula, one could express the terms $\sim b^{D,F}$ by the
octet magnetic moment. This again differs from the expression one derives from
eq.(\ref{GEs}) by terms of higher order. 
Given the rather sizeable
uncertainty of the present data, we refrain from discussing these 
differences here. Clearly, the last term in eq.(\ref{GEs}) is nothing
but the (strange) Foldy term.

\medskip

\noindent
For completeness we also give the strange magnetic form factor found in~\cite{hms}
\begin{eqnarray}
G_{M}^{(s)}(Q^2)&=&\mu_{N}^{(s)}+\frac{\pi m M_K}{(4\pi
  F_{\phi})^2}\;\frac{2}{3}\left(
                   5 D^2-6 D F+9 F^2 \right) \nonumber \\
                & &\times\left[\frac{4 M_{K}^2+Q^2}{4 M_K\sqrt{Q^2}}
                   \arctan\left(\frac{\sqrt{Q^2}}{2 M_K}\right)-\frac{1}{2}
                   \right] , \label{eq:gms} 
\end{eqnarray}
where we have introduced the strange magnetic moment of the nucleon 
\begin{eqnarray}
\mu_{N}^{(s)}&=& b_0 +\frac{1}{3}b_{}^D - b_{}^F \nonumber \\
             & &+ \frac{m M_K}{24\pi F_{\phi}^2}\left(5 D^2-6
                D F+9 F^2 \right)~.
\label{eq:ks}
\end{eqnarray}
We remark that to the order we are working, the strange form factors
are identical for the proton and the neutron. This is expected since
symmetry breaking only sets in at second order and thus should only show up
in a complete fourth order calculation.

\medskip

\noindent To summarize this section, we have given explicit expressions
for the strange (Sachs) form factors of the nucleon comprising the
various contributions from tree and one--loop graphs. To third order in
small momenta, there appear four octet and two singlet LECs. 
This has been observed  before~\cite{mus}. The octet
LECs can be fixed from standard electromagnetic nucleon and hyperon
properties as detailed in ref.\cite{khm}. The two singlet LECs play
very different roles. One of them enters directly the strange electric
radius ($d_0^{102}$), the other one ($b_0$) can be fixed from the
strange magnetic moment of the nucleon. This is the reason why to this
order the $Q^2$--dependence of the strange magnetic form factor could
be predicted without unknown parameters in~\cite{hms}. It is obvious 
that the two results from SAMPLE and HAPPEX are sufficient to pin down
the singlet LECs (within some ranges due to the presently large experimental
uncertainties).

%%%%%%%%%%%%%%%%%%%%%%%%%%%%%%%%%%%%%%%%%%%%%%%%%%%%%%%%%%%%%%%%%%%%%%%%
\section{Results and discussion}
 
We are now in the position to determine the various LECs and consequently
the strange form factors of the nucleon. To deal with the systematic,
statistical and theoretical errors given by the SAMPLE and HAPPEX collaborations,
we add these in quadrature and thus use
\begin{eqnarray}\label{Sunc}
G_{\rm SAMPLE}^{(s)}(Q_S^2) &=& 0.23\pm 0.44~, \\
G_{\rm HAPPEX}^{(s)}(Q_H^2) &=& 0.023 \pm 0.048~.
\end{eqnarray}
Together with the LECs $b^{D,F}$ and $d^{101,102}$ fixed from the proton
and neutron magnetic moments and charge radii, respectively~\cite{khm},
\begin{eqnarray}
&& b^D = 3.92~, \quad b^F = 2.92~, \nonumber \\
&& d^{101}(1~{\rm GeV}) = -1.06~, \nonumber \\
&& d^{102}(1~{\rm GeV}) = 1.70~, 
\end{eqnarray}
we can easily deduce the LECs $b_0$ and $d_0^{102}$ (assuming that
we can use the third order chiral expansion at the momentum of the HAPPEX 
experiment, see the first footnote),
\begin{equation}\label{SLECs}
b_0 = 0.06 \pm 0.44~, \quad d_0^{102} = -2.20 \pm 0.20~,
\end{equation}
leading to the singlet magnetic moment and electric radius of
$\kappa_N^{(0)} = 0.16$ and $\langle r_{0,E}^2 \rangle =
1.96\,$fm$^2$.\footnote{Note that this value appears unnaturally large
  due to our normalization of the singlet current. For the more
  conventional normalization to the baryon number, it would have to be
  divided by a factor of three.}
We remark that the value for $d_0^{102}$ is of natural size, i.e. of order one,
and that the uncertainty reflects only the experimental 
errors. For $b_0$, the central value appears somewhat small but it can
be considered natural within its sizeable uncertainty. With these
numbers, we can now evaluate the strange form factors. In what follows,
we will always give a central value (cv) based on the central values
of $b_0$ and $d_0^{102}$ and a range, which are the lower and upper bounds
we can get from combining the uncertainties $\pm \delta b_0$ and
$\pm \delta d_0^{102}$ in all possible ways (for the electric form factor). 
We consider this a conservative 
estimate of the theoretical uncertainty within the accuracy of the calculation 
presented here. It does in no way reflect an estimate about the possible 
accuracy when one goes to higher order in the chiral expansion. Such an error
is difficult to estimate since at present only very few systematic studies
in three flavor baryon CHPT exist (in the sense that all possible terms at
a given order have been retained and that the counter terms can be fixed
without any modeling.  For a recent review, see~\cite{ulfcd}).

\begin{figure}[h]
%\vspace{-10cm}
\centerline{\epsfig{file=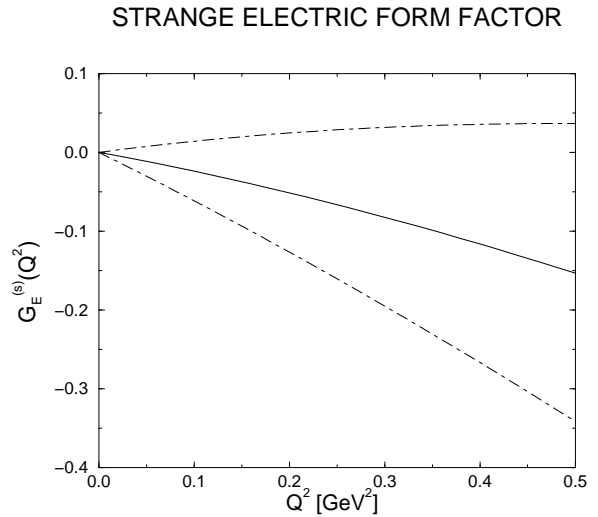,width=3in}}

\vspace{0.3cm}

\caption[diag]{\protect \small
Electric strangeness form factor of the nucleon. The solid band gives the
prediction based on the central values of the LECs and the dot--dashed
lines reflect the possible (conservative) range due to the uncertainties.
\label{fig:ges}}
\end{figure}

\medskip
\noindent
Consider first the strange electric form factor. It is shown in fig.~\ref{fig:ges}
for the central values of the LECs (solid line) and the band displayed by the
dot--dashed lines gives the theoretical uncertainty as explained above.
We remark again that this band is presumably too wide, i.e. if one were to
perform an analysis based on correlated uncertainties, this band would
shrink. 
\noindent 
We remark that these uncertainties are dominated by the uncertainty in
$d_0^{102}$, whereas the error in $b_0$ leads only to moderate changes.
This means that the contribution from the Foldy term to the strange 
electric form factor
is of much less importance as e.g. in the case of the neutron charge form
factor. From the form factor we readily deduce the strange electric
radius as defined in eq.(\ref{defrad}). We find
\begin{equation}
\langle r_{E,s}^2\rangle = (0.05 \pm 0.09)~{\rm fm}^2~,
\end{equation}
which is a fairly small and {\it positive} number, and even given the 
sizeable uncertainty, is on the lower side 
of predictions based on dispersive approaches including 
maximal OZI violation~\cite{bob,hmd}. It is more compatible with models
that include $\pi \rho$~\cite{mmsvo} or $\bar{K}K$~\cite{hm} continuum 
contributions in the isoscalar spectral functions besides the vector
meson poles ($\omega, \phi, \ldots$). Furthermore, we remark that the central value
for the strange electric radius agrees in size but not in sign with the
quark model calculation of ref.\cite{gi}. Note also that from the octet
current the strange electric radius inherits the  chiral singularity
$\sim \ln(M_K)$, cf. eq.(\ref{GEs}). The corresponding octet radius is
$\langle r_{8,E}^2 \rangle = 1.04\,$fm$^2$. It is also worth to point
out that the momentum dependence of the strange electric form factor
is rather different from the one of the neutron charge form factor,
which also vanishes at zero momentum transfer.
\medskip

\begin{figure}[h]
%\vspace{-10cm}
\centerline{\epsfig{file=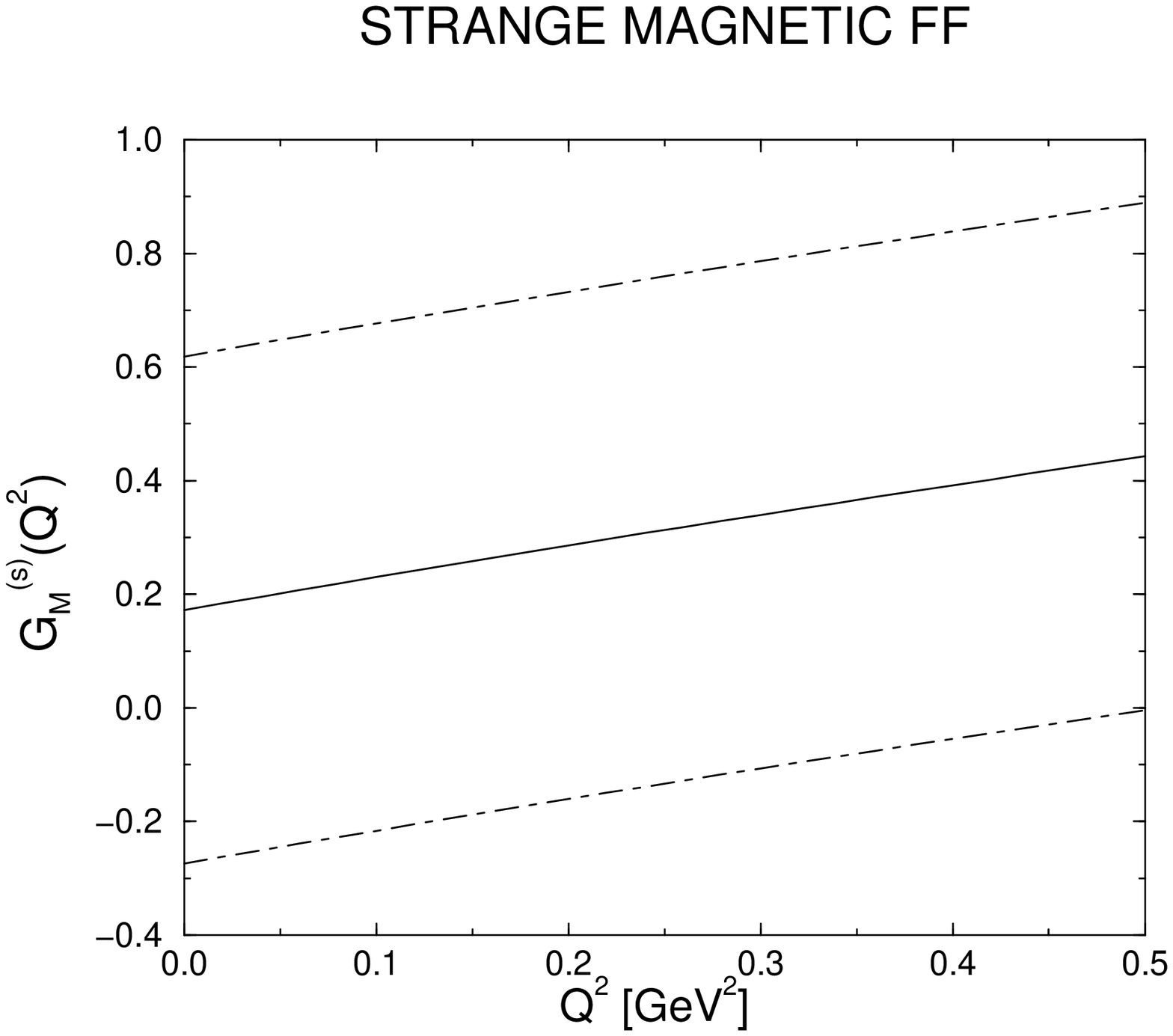,width=3in}}

\vspace{0.3cm}

\caption[func]{\protect \small
The strange magnetic form factor. For notations, see figure~1.\label{fig:gms}}
\end{figure}

\noindent We now turn to the strange magnetic form factor. Its momentum
dependence was already discussed in ref.\cite{hms}, but having fixed the
LEC $b_0$ within a certain range here, we now have an absolute prediction
for $G_M^{(s)} (Q^2)$. This is shown in fig.\ref{fig:gms}.
The rather wide band shown in fig.~2 reflects the sizeable uncertainty of
the SAMPLE result. The central value of the so determined strange magnetic moment
is, however, positive~\cite{hms} 
\begin{equation}
\mu_N^{(s, {\rm cv})} = 0.18~,
\end{equation}
and is thus at odds with most model calculations (see
e.g. table~1 in~\cite{beise}).
If one uses the relation of ref.\cite{hms} that relates the momentum dependence
of the strange magnetic form factor to the one of the isoscalar magnetic
nucleon form factor, the deduced strange magnetic moment would still be positive
but very close to zero. This shows that there is still some room for
improving the theoretical description of the magnetic form factor.
The magnetic radius is uniquely fixed in terms of well--known low energy 
parameters~\cite{hms},
\begin{eqnarray}
\langle r_{M,s}^2\rangle &=& -\frac{\pi m}{(4\pi F_\phi)^2 M_K}\frac{1}{3}
(5D^2-6DF+9F^2) \nonumber \\ 
&=& -0.14\,\mbox{fm}^2~. \label{eq:rho}
\end{eqnarray}
The slope is identical for a proton or a neutron target, 
it is {\it negative} and to this order  independent of the 
strange magnetic moment $\mu_{N}^{(s)}$.
The radius has the very reasonable behavior that in the
limit of very heavy kaons $M_K\rightarrow\infty$ it goes to zero, whereas it
explodes in the chiral limit $M_K\rightarrow 0$. 

\medskip

\begin{figure}[htb]
%\vspace{-10cm}
\centerline{\epsfig{file=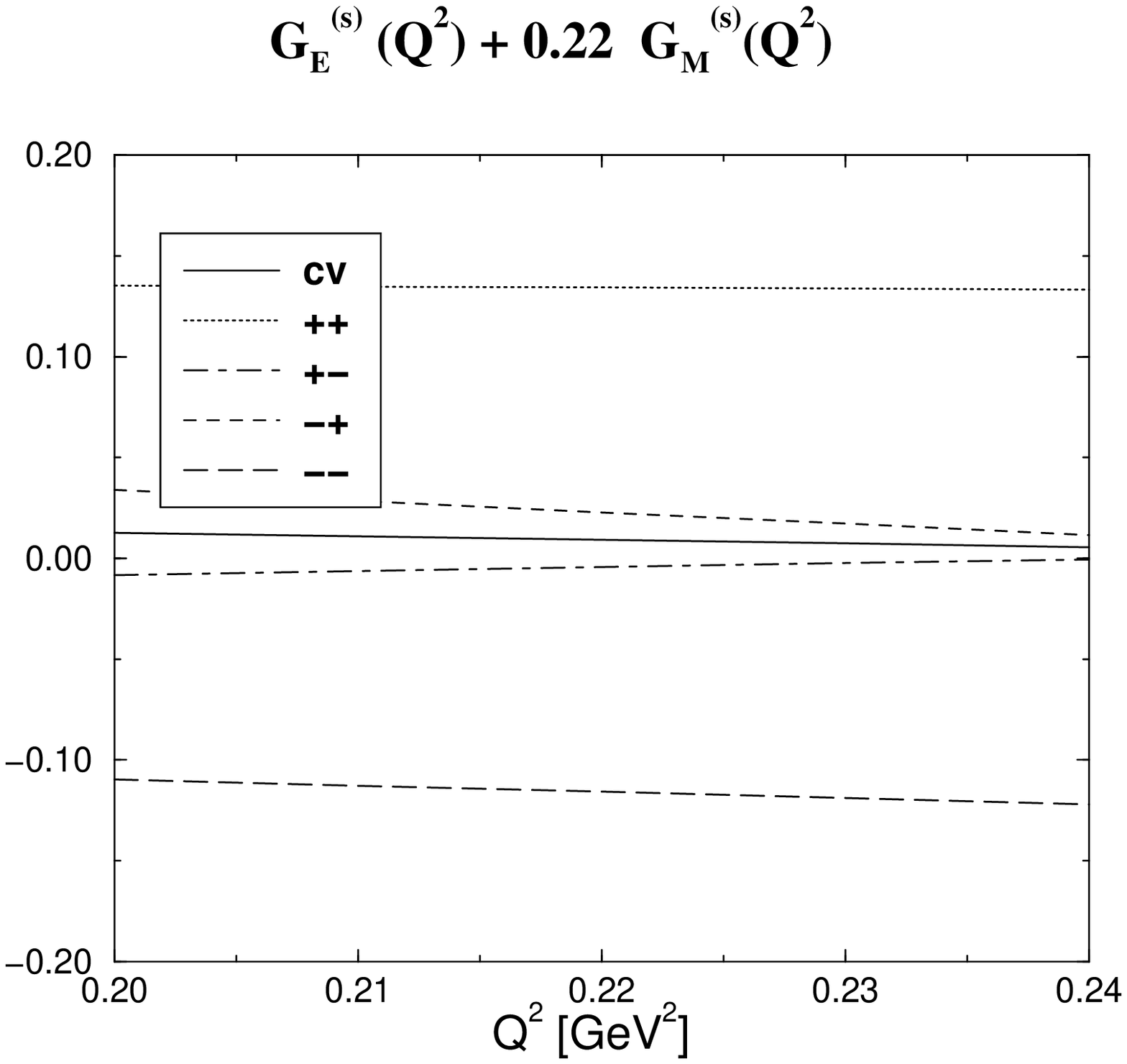,width=3in}}
%\vspace{0.3cm}
\caption[GMs]{\protect \small
$G_E^{(s)} (Q^2) + 0.22 G_M^{(s)}(Q^2)$ for momentum transfer
squared between 0.20 and 0.24~GeV$^2$. The solid line refers to the
central values of the LECs $b_0$, $d_0^{102}$  and the other
lines (dotted, dashed, $\ldots$) to the various combinations of the LECs
within their uncertainties as given in the inset (here, the first ``$\pm$''
refers to $\pm \delta d_0^{102}$ and the second to $\pm \delta b_0$, see
also table~I).
\label{fig:mami}}
\end{figure}

\noindent We now turn to the MAMI experiment, which attempts to measure
$G_{\rm MAMI}^{(s)} (Q_M^2) =G_E^{(s)}(Q_M^2) + 0.22\;G_M^{(s)}(Q_M^2)$
at a four-momentum transfer (squared) of $Q^2_M=0.23~{\rm GeV}^2$. This
value of $Q^2$ is much better suited for the chiral expansion. We find,
however, that at this value of the momentum transfer, there are sizeable
cancellations between the electric and the magnetic contributions. 
The prediction for the various combinations of the singlet LECs are 
given in table~\ref{tab1}.
\begin{table}[h]
\caption[tabmami]{Predictions for the combination of strange form
  factors to be measured at MAMI by the A4 collaboration. The central
  values of the singlet LECs are denoted by the ``*''.}\label{tab1}
\medskip
\renewcommand{\arraystretch}{1.2}
\begin{tabular}{lcr}
 $d_0^{102}$ & $b_0$ &  $G_{\rm MAMI}^{(s)} (Q_M^2)$\\ \hline
 -2.20$^*$    & 0.06$^*$ & $0.007$\\ %\hline
 -2.00        &    0.50  & $ 0.134$\\
 -2.00       & $-$0.38  & $-0.002$\\
 -2.40        &    0.50  & $0.017$\\
 -2.40        & $-$0.38  & $-0.119$ 
\end{tabular}
\end{table}
%\end{align}
\noindent
The corresponding results for a small $Q^2$ interval
($0.20 \le Q^2 \le 0.24\,$GeV$^2$) are shown in
fig.\ref{fig:mami}. Here, the uncertainty band is 
given by almost equal shares from the uncertainty in $b_0$ and the
one in $d_0^{102}$.

%%%%%%%%%%%%%%%%%%%%%%%%%%%%%%%%%%%%%%%%%%%%%%%%%%%%%%%%%%%%%%%%%%%%%%%%
\section{Summary and conclusions}

We have calculated the form factors of the strange vector current of 
the nucleon in the framework of chiral perturbation theory, updating
the analysis of ref.~\cite{mus}. To third
order in the chiral expansion, there appear six low--energy constants.
Four of these can be trivially deduced form the neutron and proton
charge radii and magnetic moments. The remaining two singlet couplings
can be determined from the recent SAMPLE and HAPPEX measurements of
combinations of the strange form factors. The crucial {\it assumption} here
is that we can apply the chiral expansion at a momentum transfer as
large as the one in the HAPPEX experiment, i.e. at $Q^2 =
0.48\,$GeV$^2$. With this cautionary remark in mind, the pertinent results of
our study can be summarized as follows:
\begin{enumerate}
\item[$\circ$] The singlet LECs given in eq.(\ref{SLECs}) are of natural
  size. The error given reflects the sizeable uncertainty of the
  experimental values obtained by SAMPLE and HAPPEX. To obtain 
  theoretical uncertainties of the LECs, we have added the various
  experimental errors in quadrature.
\item[$\circ$] For the central values of the LECs, the strange
  electric form factor of the nucleon is negative,
  cf. fig.~\ref{fig:ges}. The band given in the figure reflects the
  worst case scenario of combining the uncertainties in the singlet
  LECs (i.e. an analysis based on correlated errors would give a
  smaller uncertainty). 
  To this order in the chiral expansion, the proton and neutron
  strange electric form factor are equal with small and {\it positive}
  radius, $\langle r^2_{E,s}\rangle = (0.05\pm 0.09)$~fm$^2$.
\item[$\circ$] The strange magnetic form factor was already discussed
  in detail in ref.\cite{hms}. In fig.~\ref{fig:gms} we show the
  absolute prediction based on input from the SAMPLE result. The
  corresponding central value of the strange magnetic moment is
  $\mu_N^{(s)} = 0.18$ with an uncertainty as given in
  eq.(\ref{Sunc}). The corresponding strange magnetic radius is given
  entirely in terms of well--known parameters,  $\langle r^2_{M,s}\rangle = 
  -0.14$~fm$^2$.
\item[$\circ$] The predictions for the MAMI A4 experiment, which
  intends to measure $G_E^{(s)} + 0.22G_M^{(s)}$ at $Q^2 =
  0.23$~GeV$^2$, are collected in table~\ref{tab1}. For the central
  values of the LECs, the resulting number is fairly small due to
  cancellations between the electric and magnetic contributions.
  Due to these cancellations, varying the LECs within their
  uncertainties does not allow for a precise prediction. 
\end{enumerate}
We have shown that heavy baryon chiral perturbation theory can indeed
be used to analyze the strange form factors of the nucleon. Our study
should be considered exploratory due to the fairly large momentum
transfer involved in the HAPPEX experiment. However, with the
on--going activities at BATES, Jefferson Lab and MAMI we should soon have
an improved data base which will allow to make better use of the chiral
symmetry constraints for the strangeness vector current matrix
elements in the nucleon. Higher order calculations (possibly involving
the decuplet) are also needed~\cite{mupu}.

\bigskip

\section*{Acknowledgements}

We thank Nathan Isgur for a useful conversation.

\vfill \eject

%\vspace{-0.2cm}
%%%%%%%%%%%%%%%%%%%%%%%%%%%%%%% refs %%%%%%%%%%%%%%%%%%%%%%%%%%%%%%%%%%%%%

\vfill

\end{document}